# Immune Evasion through Competitive Inhibition: the Shielding Effect of non-Stem Cancer Cells

Irina Kareva[1]

[1]Newman Lakka Institute, Floating Hospital for Children, Tufts Medical Center,

75 Kneeland St., Boston, MA, 02111

Email: Irina.Kareva@tufts.edu

**Abstract**

It has been recently proposed that the two "emerging" hallmarks of cancer, namely altered glucose metabolism and immune evasion, may in fact be fundamentally linked (1). This connection comes from up-regulation of glycolysis by tumor cells, which can lead to active competition for resources in the tumor microenvironment between tumor and immune cells. Here it is further proposed that cancer stem cells (CSCs) can circumvent the anti-tumor immune response by creating a 'protective shield' of non-stem cancer cells around them. This shield can protect the CSCs both by creating a physical barrier between them and cytotoxic lymphocytes (CTLs), and by promoting competition for the common resources, such as glucose, between non-stem cancer cells and CTLs. The implications of this hypothesis are investigate using an agent-based model, leading to a prediction that relative CSC to non-CSC ratio will vary with the strength of the host immune response, with the highest occurring at an intermediate state of immune activation. A discussion of possible therapeutic approaches concludes the paper, suggesting that a chemotherapeutic regimen consisting of regular pulsed doses, i.e., metronomic chemotherapy, would yield the best clinical outcome by allowing CTLs to most effectively reach and eliminate CSCs.

**Introduction**

The immune system is generally very effective in preventing budding tumors from progressing to a fully malignant disease (2-4). However, as the cytotoxic cells of the immune system attempt to eliminate the tumor, they effectively select for non-immunogenic clones, thus delaying but not arresting tumor progression, a process that has become known as tumor immunoediting (2; 5-7). Moreover, tumors that have been 'edited' in this fashion are typically more aggressive than their 'unedited' counterparts due to selection for more immunoresistant clones; a property first demonstrated by Shankaran and colleagues (8).

It has been recently proposed that the process of immunoediting may additionally be driven by competition for resources between cancer and immune cells in the tumor microenvironment (1). Specifically, the following microenvironmentally-based immunoediting scenario has been proposed:

1) As the tumor increases in size while not yet recognized by the cytotoxic cells of the immune system, it begins to outgrow its blood supply, causing the formation of oxygen-deprived regions in the tumor interior.

2) Intra-tumoral hypoxia causes the affected tumor cells to turn to glycolysis as a primary mode of glucose metabolism; an event known to be accompanied by an up to 30-fold up-regulation of the activity of glucose transporters (9).

3) Once activated, the CTLs first attack the topologically accessible outer rim of the tumor, what we term the 'protective shield', thus effectively exposing the tumor's 'glycolytic core'.

4) The exposed glycolytic tumor cells, which have been forced to up-regulate their glucose transporters, now actively compete with CTLs for glucose.

5) In a nutrient-deprived state, CTLs are less able to perform their tumoricidal function, allowing the tumor as a whole to evade immune attack.

In this way, the process of immune evasion may be the result of a competition for resources in the tumor microenvironment.

Here, this construct is expanded to incorporate cancer stem cells (CSCs). Because CSCs comprise a relatively small fraction of the tumor population on account of their high probability of asymmetric division (producing one CSC and one non-stem cell) (10), they will tend to surround themselves with non-stem progeny, which then could follow the aforementioned immunoediting scenario . The non-CSCs will then protect CSCs from cytotoxic lymphocytes both by creating a physical barrier and through outcompeting CTLs for essential nutrients. These considerations are summarized in Figure 1.

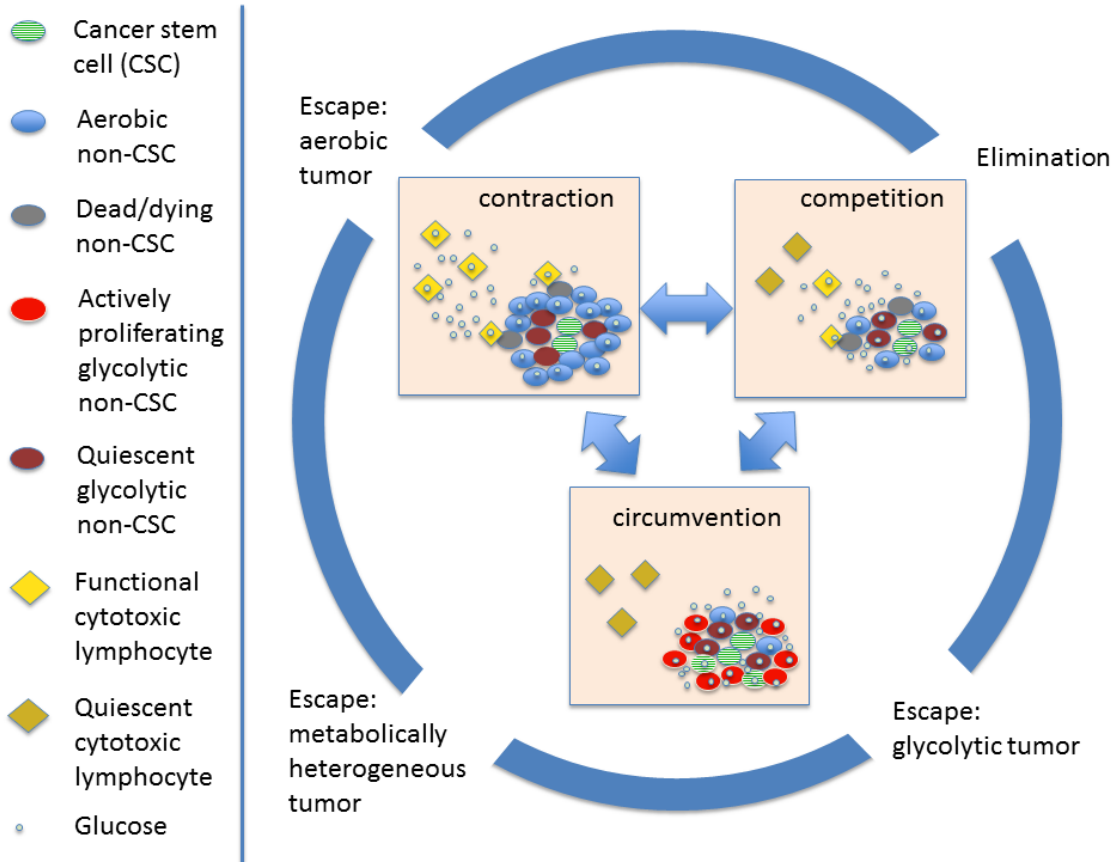

Figure 1. Schematic representation of the proposed immunoediting construct, which incorporates CSCs using aerobic and particularly glycolytic non-CSCs as a protective 'shield' from cytotoxic lymphocytes.

The proposed theoretical framework allows making predictions about CSC content in different tumors with regards to the immune response, which can be of vital importance, as CSCs, the driving cells in cancer progression, are believed to be more resistant to cytotoxic therapy than non-CSCs (11-14). In this case, if CSCs produced too few non-CSCs, protection from the immune system could be insufficient. If CSCs produced too many non-CSCs, then the progeny may themselves become limiting to CSC expansion, offsetting their benefit as an immune shield. Successful tumors might therefore be predicted to be those that have some optimal intermediate CSC-to-progeny proportion. In addition, this proportion may well be different for different initial states of the immune system.

To test these ideas, an agent-based model is implemented, incorporating the foregoing and other basic assumptions about the properties of CSCs, non-CSCs, immune cells and their interactions with each other and with the resources in the tumor microenvironment. We use this model to illustrate the proposed immunoediting construct and in particular, to evaluate the proposition that varying proportions of CSCs to non-CSCs might be expected in successful tumors under different immune conditions.

**Model description**

Here we build upon a previously introduced agent-based model (15), adapting it to describe the interactions between cancer stem cells (CSCs), aerobic and glycolytic non-stem cancer cells (non-CSCs), and cytotoxic lymphocytes (CTLs) with each other and their microenvironment. The model is implemented in Netlogo 5.0.2, a freely available agent-based modeling platform (16). Space in this simulation is represented by a 2-dimensional 51x51 lattice. Each patch (square) in the lattice represents a discrete microenvironment, the properties of which are detailed below, for a total of 2601 microenvironments. Cells occupy coordinates in continuous space, and more than one cell can occupy a single microenvironment. Time is modeled in discrete steps.

**Parameter estimation**

Parameter estimation is a common challenge in analyzing results from computational models. One particular difficulty lies in the fact that the estimates obtained in an experimental setting are typically a result of a multitude of complex interconnected processes, which cannot always be separated out for the purposes of a conceptual proof-of-concept model. We address this issue in the following ways. A number of parameters, such as cell growth rates, were previously estimated in a model introduced and analyzed in (15), where the authors utilize an agent-based model to evaluate the impacts of disregulated glucose metabolism on overall tumor dynamics. We modify their proposed model, shifting the focus of our investigation away from evolution of cell motility and instead introducing immune cells and cancer stem cells as active competitors for glucose in the tumor microenvironment. Nevertheless, parameters that pertain to tumor growth rates and glucose consumption rates can be well translated into our model.

For a number of other parameters, while it may not always be possible to obtain exact values for reasons outlined above, we attempt to make estimations based on the parameters' relative values to each other. That is, while we may not be able to estimate exact rates of glucose consumption by cytotoxic immune cells, for instance, we can nevertheless support the notion that highly proliferative CTLs, just as glycolytic tumor cells, consume more energy than do aerobic tumor cells (17). Morover, Buttgereit and Brand (18) reported that in a model of concanavlin A-stimulated thymocytes, where it is possible to account for over 80% of ATP consumption, protein synthesis was the most sensitive to limitations in glucose supply, followed by DNA and RNA synthesis and then by ion transport across plasma membrane. Motility also has been reported to incur significant ATP costs for immune cells due to energy requirements for cytoskeleton rearrangement during movement (19). Therefore, while we may not be able to obtain the exact values for the costs of motility and cytotoxicity for CTLs, we can nevertheless assert that these costs exist, and are incorporated as such. Moreover, since we do not have exact estimates for these costs, they are taken to be a unit of energy each, an assumption which can be relaxed or modified pending data availability.

The specific assumptions for each of the cell types and for the microenvironments are as follows:

**Properties of aerobic non-CSCs**: Aerobic non-CSCs require glucose and oxygen for survival and glucose for proliferation, although they have lower glucose requirements than glycolytic non-CSCs. They die at some natural rate, as well as after a certain number of divisions, and can also be killed by CTLs. They can 'switch' to the glycolytic mode of metabolism under conditions of oxygen deprivation. When a cell divides, the daughter cell moves to a random neighboring patch.

**Properties of glycolytic non-CSCs**: Glycolytic non-CSCs do not require oxygen, but require glucose for survival and proliferation; they consume more glucose to support these processes than aerobic non-CSCs. They die at some natural rate, as well as after a certain number of divisions, and can also be killed by CTLs. They can 'switch' to the aerobic mode of glucose metabolism under conditions of sufficient oxygen availability. When a cell divides, the daughter cell moves to a random neighboring patch.

**Properties of cancer stem cells**: Depending on oxygen availability, CSCs consume oxygen and glucose like aerobic or glycolytic cells. Each CSC division occurs symmetrically with a certain probability. If a CSC divides asymmetrically, the daughter non-CSC cell is initialized to be aerobic, a state which may change. Each cell can die due to nutrient deprivation or be killed by CTLs. CSCs can divide an unlimited number of times. When a CSC divides, the daughter cell moves to a random neighboring patch.

**Properties of cytotoxic immune cells**: It is assumed that all cytotoxic CTLs rely on glycolysis for glucose metabolism, consuming nutrients at the same rate as glycolytic tumor cells. CTLs can proliferate only under conditions of sufficient glucose supply. They can die from nutrient deprivation. Unlike tumor cells, CTLs are motile and move randomly throughout the microenvironment. As motility is costly, each step to a random neighboring patch costs CTLs a unit of energy. CTLs can kill tumor cells once they encounter them; however, killing a tumor cell also costs a unit of energy. CTLs are replenished at some constant rate when the number of immune cells in the tumor microenvironment becomes too low.

**Properties of the microenvironment**: On each patch there is a certain amount of glucose and oxygen. Both glucose and oxygen are replenished differentially, depending on the number of cells on the patch. Too many cells on the patch lowers the rate of replenishment, which simulates the crowding effect. Glucose and oxygen are taken up by different cell types at different rates, depending on whether the cell type is aerobic, glycolytic, stem or immune.

It is worth noting that while alternative energy sources, such as glutamine, are available to tumor cells, it has been shown by MacIver and colleagues (20) that activated T cells are unable to perform their function in the absence of glucose, even in the presence of amply available glutamine. Therefore, glutamine was not incorporated as an energy source in the proposed model as it is not the driving force for competition between tumor and immune cells.

The model is initialized with a single CSC and a pre-set initial number of cytotoxic immune cells placed randomly throughout the lattice. At each time step, the following sequence of steps is realized:

1) Each tumor cell evaluates oxygen availability in its microenvironment, which determines the cell's subsequent mode of glucose metabolism.
2) Each cell consumes glucose, subtracting differentially from the glucose available on the patch and adding a number of energy units to the cell's storage, according to its 'metabolic phenotype', i.e., aerobic or glycolytic; glycolytic cells consume glucose, while aerobic cells consume both glucose and oxygen.
3) Cytotoxic lymphocytes move randomly throughout the microenvironment, killing the tumor cells that they encounter; both motility and the process of cell killing have a cost of a unit of energy each; CTLs die if their energy stores are depleted and cannot be replenished.
4) Each cell reproduces with a certain fixed probability, given adequate resources in the microenvironment; the daughter cell moves to a random neighboring patch; if the cell is a CSC, it divides symmetrically or asymmetrically with a certain probability; in case of asymmetric division, the daughter cell is initialized as aerobic non-CSC.

5) Glucose and oxygen are replenished in the microenvironment depending on the number of cells on each patch, with a higher rate of replenishment in the areas of lower cell density.
6) CTLs are replenished by the value of parameter $i_0$ if the number of immune cells on the lattice falls below a critical threshold.

Since the purpose of the proposed model is to offer a 'proof-of-concept' as to whether tumor composition can indeed be affected by the initial state of the immune system of the host, parameter values were chosen as relative to each other rather than matching them to fit a particular data set.

It is hypothesized that in successfully formed tumors, the relative proportion of non-CSCs will increase proportionally to the strength of the immune response because CSCs will require the shielding effect of non-CSCs to protect them from CTLs. Specifically, an increase in the proportion of glycolytic non-CSCs is predicted, since they are capable of competing most effectively with CTLs for glucose in the tumor microenvironment.

**Results**

The numerical experiments were designed in such a way as to enable evaluation of tumor composition at each time point with respect to the state of the immune system. The initial state of the immune response was modeled with parameter $i_0$=0, 10, 20, 30, 35, 40. For each $i_0$, 20 simulations were run for 1000 time steps, and the mean number of CSCs, aerobic and glycolytic non-CSCs were reported; only the data for tumors that had at least one CSC at t=1000 were collected. The data is represented in two types of plots: the change over time of tumor composition for each respective $i_0$ and the final ratio of CSC to non-CSC at t=1000 for each $i_0$.

**Proportion of glycolytic cells increases with the strength of the immune response**

Our simulations show that indeed, stronger immune response predicts larger overall proportion of glycolytic non-CSCs in the tumor (see Figure 2), which within the proposed theoretical framework is explained by the increased 'need' of CSCs to protect themselves from the immune system. However, the largest proportion of CSCs in the tumor at t=1000 was observed at the intermediate values of $i_0$ (see Figure 3).

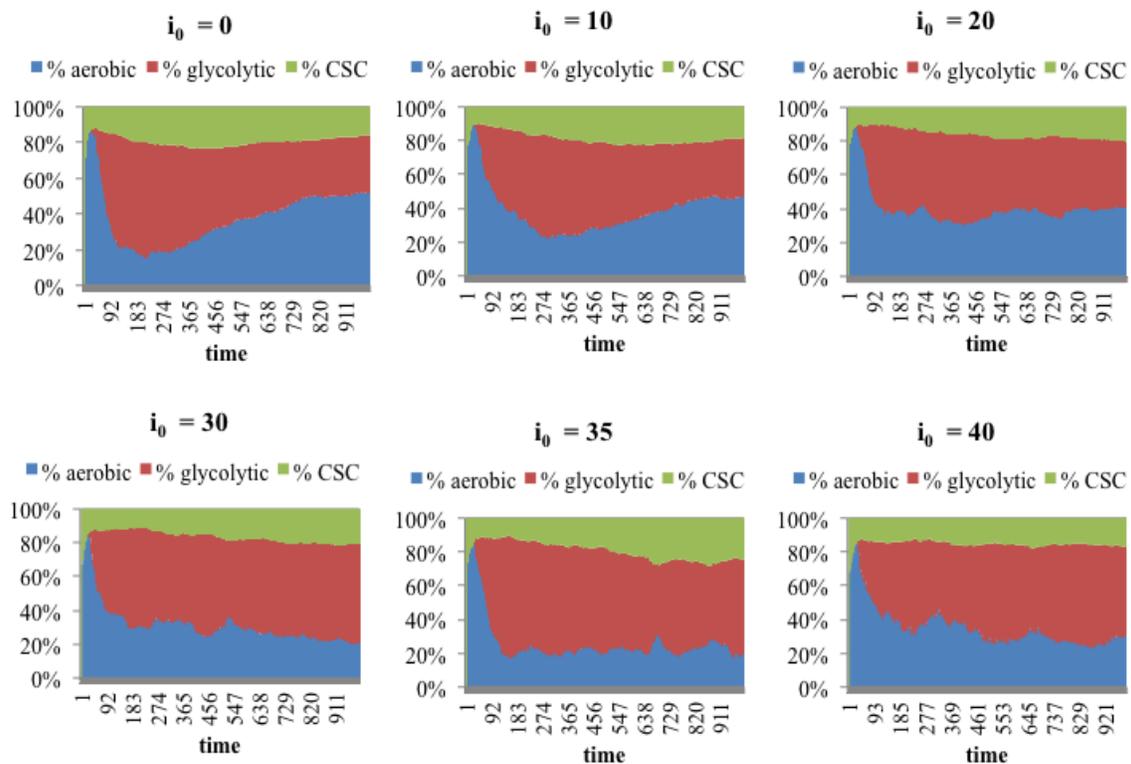

Figure 2. Change in tumor composition of CSCs (green) to glycolytic nCSCs (red) to aerobic nCSCs (blue) over time for varying initial states of the host immune system, represented by parameter $i_0$. As one can see, the proportion of glycolytic nCSCs tends to increase with $i_0$ as CSCs require increased protection from CTLs. Threshold of sensitivity to oxygen for all cells is set to $\omega=5$.

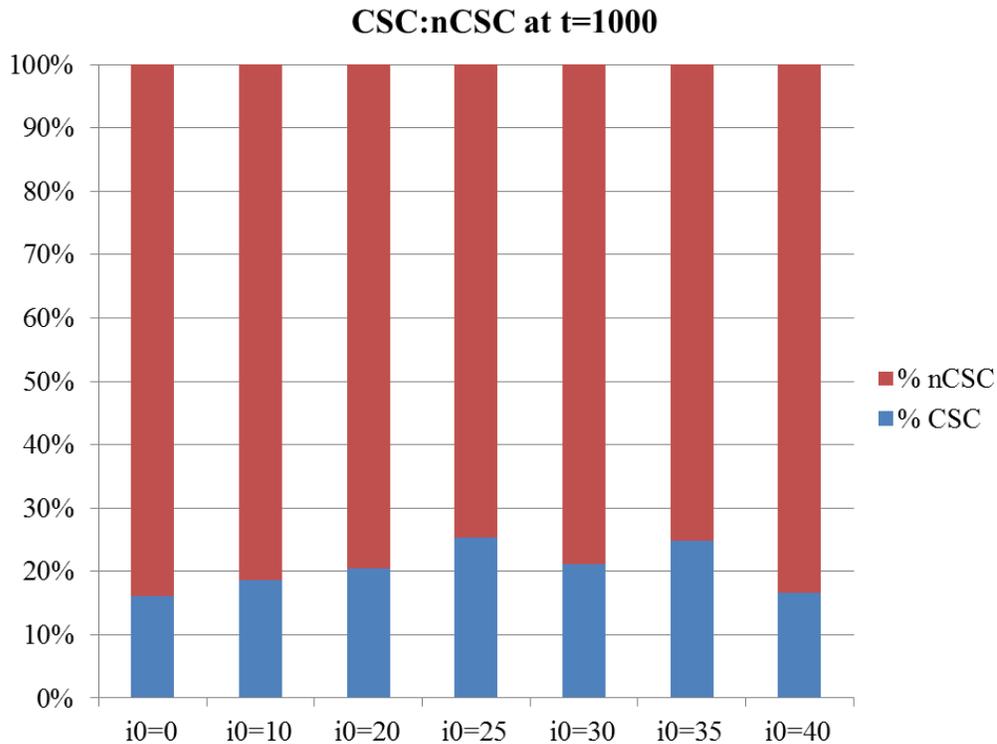

Figure 3. Proportion of CSCs and nCSCs at t=1000 for all varying initial states of the host immune system, represented by parameter $i_0$. As one can see, the highest proportion of CSCs is observed not for immunodeficient or extremely immunocompetent host, but for intermediate values of $i_0$. Threshold of sensitivity to oxygen for all cells is set to $\omega=5$.

The following explanation for this phenomenon is proposed:

At low values of $i_0$, aerobic non-CSCs that are located on the tumor rim have the greatest competitive advantage compared to other tumor cells, as they have the most access to oxygen and glucose compared to both CSCs and glycolytic non-CSCs inside the tumor. Therefore, since they are not 'sacrificed' to the immune system, aerobic non-CSCs outcompete CSCs.

At high values of $i_0$, the tumor is not able to grow to a large enough size to create sufficiently hypoxic regions that would allow for the appearance of large clusters of glycolytic non-CSCs that would present sufficient competition to cytotoxic lymphocytes. In this case, the non-CSC 'shield' would not be strong enough to contain CTLs.

At "intermediate" values of $i_0$, the tumor can grow to be large enough in size to have sufficiently large hypoxic regions to require up-regulation of purely glycolytic mode of glucose metabolism in a sufficiently large number of non-CSCs, while the immune system removes enough of the aerobic non-CSCs at the outer rim of the tumor to prevent them from outcompeting the rest of the tumor cells.

Therefore, one should expect the largest proportions of CSC in hosts with "intermediate" states of immune system. (Noticeably, these observations can be made based on assumptions of relative values of parameters, without requiring exact parameter estimation from specific data sets).

It has been reported by some groups that CSCs may have lower immunogenicity compared to non-CSCs (21,22). Intriguingly, while the same pattern of results was observed for various modifications in parameter values for costs of CTL motility and cytotoxicity (data not reported), differences were observed when the probability of recognition of CSCs by CTLs was decreased by 50% (Figures 4 and 5); this value was chosen arbitrarily due to lack of specific data. It appears that the proportion of CSCs varies with respect to the values of $i_0$ only if CSCs are as likely to be killed by CTLs upon encounter as are non-CSCs. Otherwise, proportion of CSC:non-CSC remains the same regardless of the strength of the immune response.

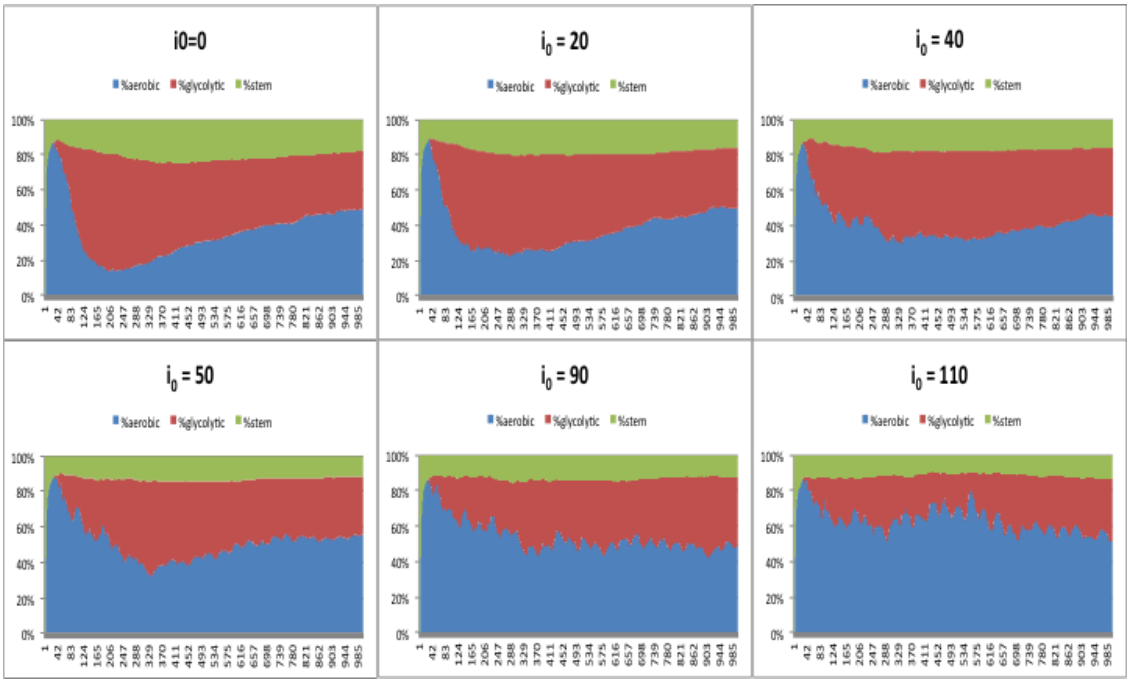

Figure 4. Change in tumor composition of CSCs (green) to glycolytic nCSCs (red) to aerobic nCSCs (blue) over time for varying initial states of the host immune system, represented by parameter $i_0$. The probability of recognition of CSC by a CTL is 50% lower than probability of recognition of nCSC. Threshold of sensitivity to oxygen for all cells is set to $\omega=5$.

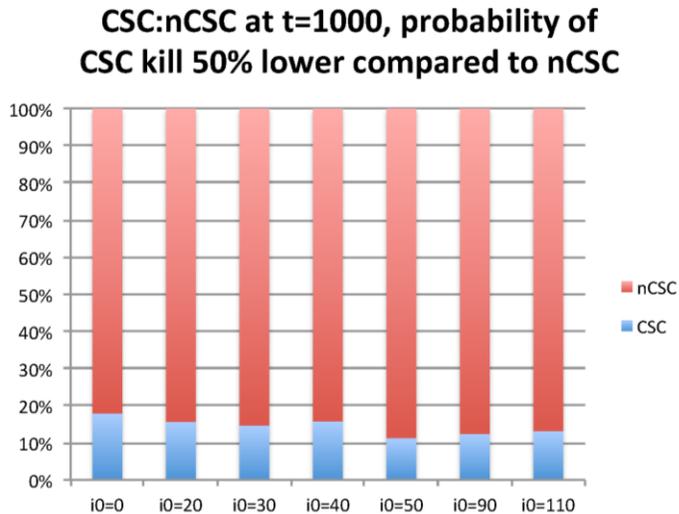

Figure 5. Proportion of CSCs and nCSCs at t=1000 for all varying initial states of the host immune system, represented by parameter $i_0$. The probability of recognition of CSC by a CTL is 50% lower than probability of recognition of nCSC. Threshold of sensitivity to oxygen for all cells is set to $\omega=5$.

**Variability in sensitivity to oxygen does not affect final tumor composition regardless of the state of the immune system**

In the previous set of *in silico* experiments, it was assumed that all cells have a fixed threshold of sensitivity to oxygen and therefore revert to either aerobic or glycolytic mode of glucose metabolism based on a criterion that is uniform for all cells. It is however possible that if there were heterogeneity among cells based on this metric, one could observe selection for cells that are either more or less sensitive to oxygen concentrations in their microenvironment depending on the state of the immune response. That is, one could observe selection towards lower oxygen sensitivity thresholds (described by the value of parameter $\omega$) under the conditions of increased selective pressure from the immune system.

In order to evaluate this hypothesis, the above-described protocol was modified to introduce mutations in parameter $\omega$ every time a tumor cell divides. The changes in the mean value of $\omega$ were recorded for all three cell types for all $i_0$. The overall tumor composition changed in the same way with respect to different $i_0$ as it did for the case of constant $\omega$ (Figure 6), and the largest proportion of CSCs was also observed for the intermediate values of $i_0$ (Figure 7). The mean value $\omega$ for glycolytic cancer cells did oscillate increasingly with stronger immune response (Figure 8). However, a 2-tailed type 1 t-test did not reveal any statistically significant changes either in overall tumor composition or in the values of $\omega$ at $t=1000$ for either of the three tumor cell types (Figure 9), These results suggest that it is not oxygen availability but competition for glucose, both within the tumor and between cancer and immune cells, that is the driving force behind the observed dynamics of immune evasion.

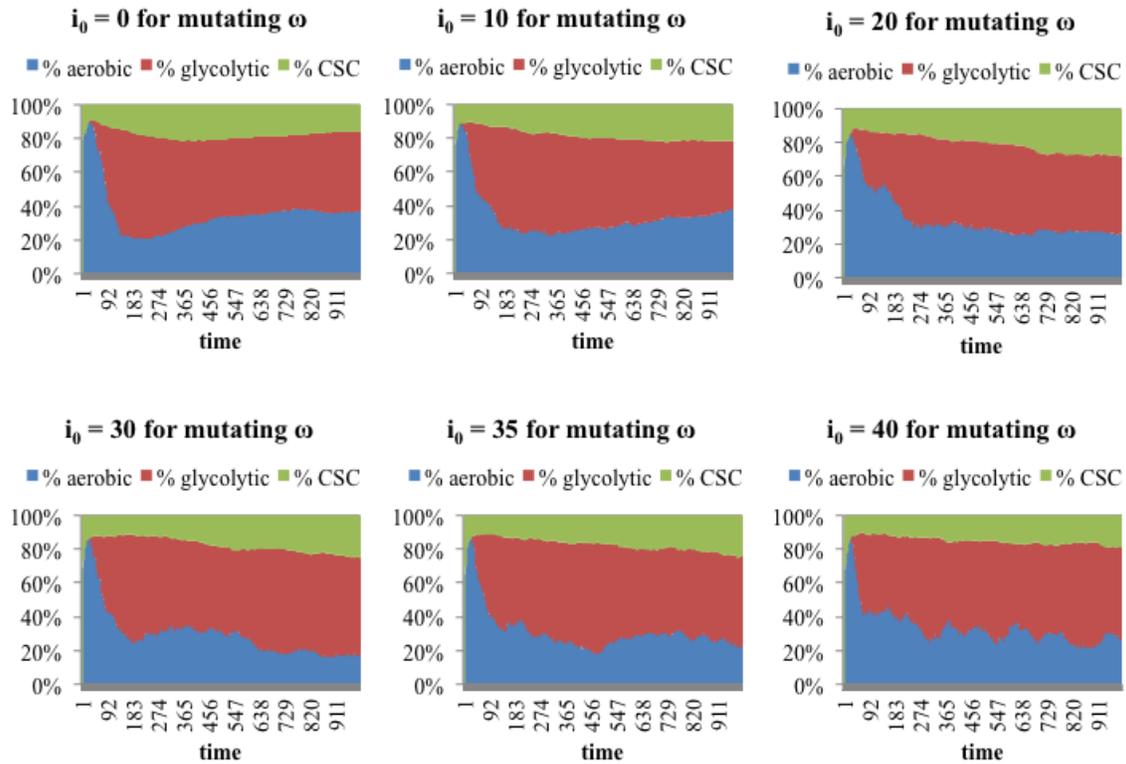

Figure 6. Change in tumor composition of CSCs (green) to glycolytic nCSCs (red) to aerobic nCSCs (blue) over time for varying initial states of the host immune system, represented by parameter $i_0$. As one can see, the proportion of glycolytic nCSCs tends to increase with $i_0$ as CSCs require increased protection from CTLs. Threshold of sensitivity to oxygen for all cells is initiated at $\omega=5$ but is allowed to mutate with each cell division for each cell.

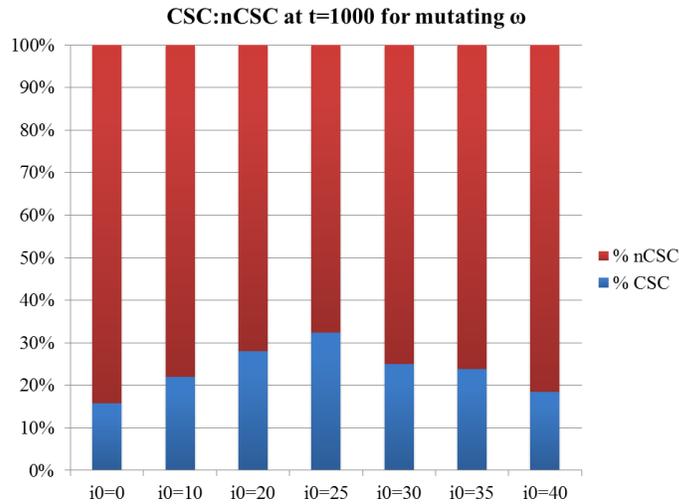

Figure 7. Proportion of CSCs and nCSCs at t=1000 for all varying initial states of the host immune system, represented by parameter $i_0$. As one can see, the highest proportion of CSCs is observed not for immunodeficient or extremely immunocompetent host, but for intermediate values of $i_0$. Threshold of sensitivity to oxygen for all cells is initiated at $\omega=5$ but is allowed to mutate with each cell division for each cell.

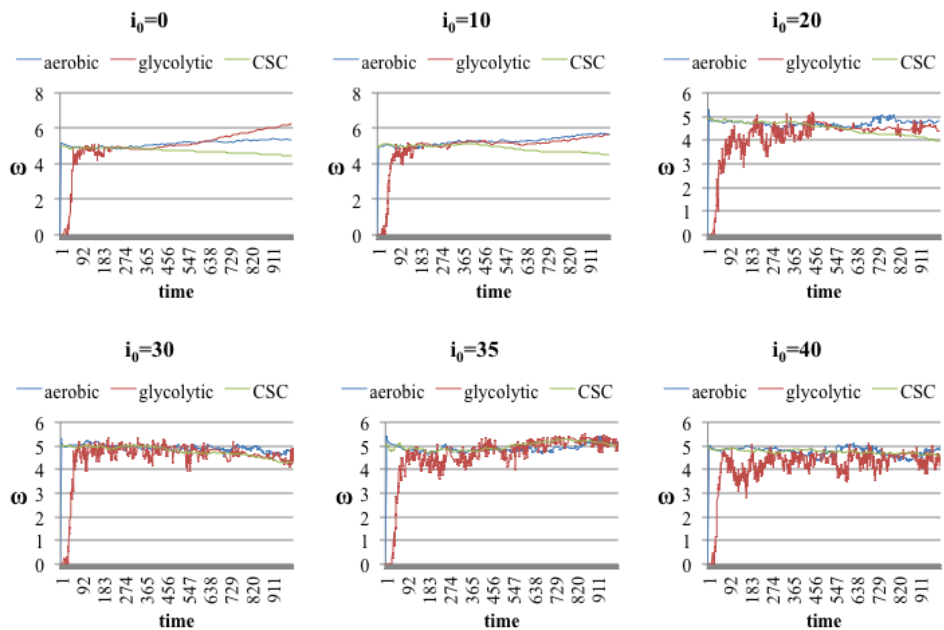

Figure 8. Change in the mutating threshold of sensitivity to oxygen (parameter $\omega$) over time for varying initial states of the host immune system, represented by parameter $i_0$. As one can see, while $\omega$ oscillates increasingly for larger $i_0$, the mean of $\omega$ remains largely the same for all three cell types.

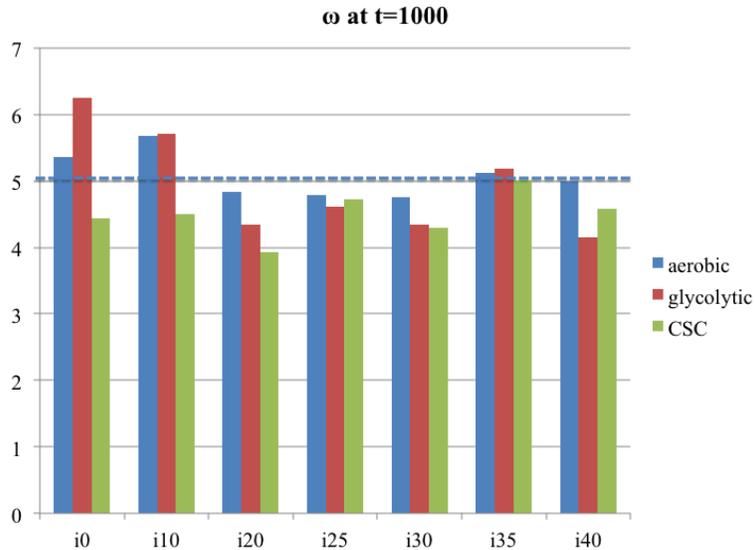

Figure 9. Mean value of the threshold of sensitivity to oxygen (parameter ω) at t=1000 for varying initial states of the host immune system, represented by parameter $i_0$, compared to fixed value of ω=5 for the first set of simulations. While there are slight changes in ω for all three cell types, none of them were statistically significant as evaluated by a 2-tailed type 1 t-test, suggesting that it is not oxygen availability.

**Discussion**

Here a theoretical framework is proposed, where CSCs can escape recognition by cytotoxic lymphocytes by creating a 'shield' of differentiated non-CSCs around them. These shielding cells not only provide a physical barrier to the immune cells but also create a local nutrient flux, thereby starving CTLs out of functionality. Specifically, we proposed that if a CSC divided asymmetrically before it was recognized by the CTLs, and if the progeny grew to form a sufficiently large tumor to cause the appearance of hypoxic regions and consequent reversion by many of the tumor cells to purely glycolytic mode of glucose metabolism, then, once the immune cells eventually clear the aerobic cells on the outer rim of the tumor, they will encounter competition for glucose from glycolytic cancer cells. Successful acquisition of the common glucose, made possible by up-regulation of nutrient transporters of glycolytic cancer cells while in the hypoxic regions of the tumor, could cause incapacitation of CTLs due to starvation, thereby allowing for continued unrestrained tumor growth (see Figure 1). Within this construct one would also expect CSCs to be found primarily in hypoxic regions of the tumors, where they would be most protected from the immune response.

In order to evaluate the proposed hypothesis,.an agent-based model was created, focusing on interactions between CSCs, aerobic and glycolytic non-CSCs with varying glucose demands, , and motile cytotoxic immune cells, whose functionality is nutrient dependent. Simulations were run for 1000 time steps, for various initial states of the immune system of the host, from complete immunodeficiency to extremely strong immune response. Tumor composition was then evaluated for all of the initial immune states.

The proportion of glycolytic non-CSCs increased dramatically with increased immune response (Figure 2), supporting the initial hypothesis. Moreover, the proportion of CSCs was largest not for the strongest or weakest states of the immune response but for intermediate ones (Figure 3), supposedly because in this case the tumor can grow large enough to create hypoxic regions but it small enough that the non-stem progeny do not outcompete CSCs.

Next, we hypothesized that if each cell had an individual threshold of sensitivity to oxygen, which would determine oxygen-based threshold for a "glycolytic switch", then over time selection towards cells with lower threshold could be observed. However, while the threshold of sensitivity to oxygen ω did change slightly over time, the changes both in tumor composition, as well as in the mean values of ω for all three cell types was not statistically significant (Figures 8 and 9), suggesting that oxygen availability is less of a crucial determinant of the outcome of tumor progression compared to glucose.

It is of course not possible to claim that competition for common nutrients is the sole mechanism whereby CSCs can avoid recognition by the immune system. Instead, we propose that competition for resources can provide an additional mechanism whereby CTLs are incapacitated, contributing to immune evasion. Noticeably, modifying the model to increase or decrease the actual costs of motility and cytotoxicity of CTLs does not affect the predicted pattern of behavior (data not reported). Similarly, modification of the level of possible infiltration of the tumor by CTLs based on crowding effects did not result in qualitative changes in dynamics (data not reported). This can be explained by the fact that even when CTLs are physically able to infiltrate the simulated tumor, they are quickly starved by glycolytic tumor cells in the hypoxic tumor core, thus achieving the same qualitative effects as artificially imposed restrictions on CTL infiltration.

The predicted pattern dynamics were robust to changes in all parameters evaluated with the exception of sensitivity of CSCs to CTLs. Decreasing the probability of cell kill for CSCs upon encounter with CTLs by 50% compared to non-CSCs resulted in a similar pattern with respect to changes in tumor composition over time for various $i_0$ (Figure 4) but the prediction about proportions of CSC:non-CSC at t=1000 was not preserved (Figure 5). Literature about sensitivity of CSCs to CTLs is inconclusive: while a number of authors report lower sensitivity of CSCs to being recognized by CTLs compared to non-CSCs (21, 22), it is unclear whether this is solely an intrinsic property of CSCs, such as lower expression of MHC-II molecules (22), or a by-product of the 'shielding' effect of non-CSCs. Other groups report efficient recognition of CSCs by CTLs (23), which suggests that this is a question that in fact pends further investigation.

One possible way to understand the interactions of CSCs with their environment is from the point of view of normal processes that occur during embryogenesis or tissue regeneration. Tissue injury often results in locally created hypoxia due to damage to blood vessels, which results in local acidosis as the cells downstream from the damaged vessels become oxygen deprived. Lower pH appears to create a locally immunosuppressive environment (1), which would allow tissues to rebuild themselves. Normalized vascularization after tissue regeneration would allow for exposure of adult stem cells to immune cells, thus providing an additional regulatory mechanism to stop generation of new tissues. Tumors in turn could harness this mechanism, continuing to grow through creating a hypoxic environment, inadvertently protecting CSCs. This hypothesis will require further investigation.

**Game theory and cancer**

It has become increasingly recognized and accepted that tumor dynamics is governed largely by the process of Darwinian selection, where tumor cells experience and respond to various selective pressures, such as competition for resources with each other and somatic cells, interaction with predators (the immune system), migration (metastases), all within the 'ecosystem' of the human body (24-31). One of the well-developed analytical approaches to understanding evolving populations is evolutionary game theory, which has already been partially applied to cancer (33-39). The main premise of this approach is the assumption that individuals, such as cells, operate for immediate optimization of their payoff, which in the context of evolutionary game theory is fitness, i.e., survival and the number of progeny. Individuals may use different strategies to achieve this goal, and understanding of these strategies and their immediate consequences can provide understanding about why a system is evolving in a particular direction.

In the context of the proposed model, we assume that it is CSCs that want to maximize their fitness and produce a maximal number of other CSCs. They can achieve this goal using two strategies: symmetric or asymmetric division, yielding either two CSCs or one CSC and one non-CSC, respectively. Symmetric division allows yields more CSCs immediately. Asymmetric division may allow for survival of the progeny in the long term.

From this point of view, in the absence of the immune response, or if it did not affect the growth dynamics CSCs, one would expect eventual CSC enrichment, as is predicted by a number of previously published theoretical models(40, 41), since it is symmetric division that would allow CSCs to maximize their fitness. However, in the presence of the immune response, the 'optimal strategy' becomes mixed, since in order to maximize their fitness CSCs now have to divide asymmetrically to protect themselves from the immune cells, and the position of the 'equilibrium strategy', i.e., the number of times the CSC would need to divide symmetrically or asymmetrically in order to maximize its fitness, would be different depending on the initial state of the immune response. Perhaps, manipulating the tumor microenvironment in such a way as to select for the increased asymmetric CSC division could be a potentially fruitful strategy to make the tumor more vulnerable to anti-cancer therapies.

**Therapeutic implications**

Cancer stem cells are hypothesized to be the driving force behind unrestrained tumor growth, as they are also resistant to a number of cytotoxic therapies (13; 42-45). One possible way to circumvent this problem could involve taking advantage of the immune cells, which might be able to eliminate CSCs, even if less efficiently compared to non-CSCs (23,46-48). However, if the proposed construct is correct, CTLs are unable to do so because of the 'shield' of differentiated tumor cells that restricts their access to CSCs. Therefore, the optimal therapeutic strategy in this case would be not the one to cause the most extensive cell mortality but the one that will facilitate the access of cytotoxic immune cells to CSCs.

Two main considerations CSCs and CTLs need to be taken into account. Firstly, both theoretical considerations, outlined here, and experimental evidence suggest that CSCs are to be primarily found in the hypoxic areas of the tumor (49, 50), where they are the most 'protected' from the immune cells. Secondly, extensive cytotoxic therapies are highly damaging to the immune cells themselves (51), and consequently gained access the tumor core and the CSCs would be ineffective if there remain no immune cells that could then attack the CSCs.

The three major approaches that can be taken in chemotherapy administration are MTD (maximum tolerated dose), metronomic, and continuous therapy, which is a particular case of the metronomic therapy.

As the name suggests, MTD involves administering the highest possible dose of cytotoxic drugs that a patient can tolerate, with the intention of damaging a maximal number of cells and expectation that normal tissues would be able to recover better than cancerous ones. The major drawback of MTD from the point of view of the considerations, outlined above, is the damage that it causes to the immune cells (51). Therefore, even if this could be a good way to reach the CSCs, it is very likely that the time it would take for the anti-tumor immune cells to recover would be greater than the time it would take for the CSCs to form another 'shield'. Noticeably, these predictions should hold as long as CSCs can be recognized by CTLs.

Continuous therapy involves administering very low doses of chemotherapy on a daily basis, ensuring constant presence of the cytotoxic drug in the patient's body. While this approach is much less toxic to the patient, it still has a problem of giving no time for the anti-tumor immune cells to recover, since they would be constantly eliminated by continuous inflow of cytotoxic chemical agents.

Metronomic therapy provides an intermediate case between MTD and continuous therapy, and involves administering lower doses of chemotherapy at more frequent intervals, but not on a daily basis. This approach could ensure continual 'peeling off' of the outer layers of the tumor, which protect the CSCs, thus giving anti-tumor immune cells access to the tumor core, while allowing sufficient time for CTLs to recover from the damaging effects of the therapy and attack the CSCs, thereby giving the best chance for more successful long-term tumor elimination.

**Conclusions**

Limited success of some therapies may be rooted not in the lack of effectiveness of the cytotoxic drugs themselves but in the lack of understanding of tumor topology. If the differentiated non-CSC cells do indeed provide a barrier to both chemotherapeutic agents and the body's natural defenses, then the problem of halting cancer progression might be overcome with appropriate mode of therapy administration. Specifically, if therapy-resistant CSCs do shield themselves from anti-tumor immune cells by 'hiding' in the tumor's hypoxic regions, then slowly exposing these areas using moderate but relatively frequently administered levels of chemotherapy might give a better chance for cytotoxic lymphocytes to eliminate the tumor.

**Acknowledgments**

The author would like to thank Philip Hahnfeldt, Lynn Hlatky and Heiko Enderling for extremely valuable discussions and suggestions during manuscript preparation. This research was supported by the Office of Science (BER), U.S. Department of Energy, under Award Number DE-SC0001434 (to Philip Hahnfeldt).

**References**

1. Kareva I, Hahnfeldt P. The Emerging "Hallmarks" of Metabolic Reprogramming and Immune Evasion: Distinct or Linked?. Cancer Res. 2013 Feb 19.
2. Dunn GP, Bruce AT, Ikeda H, Old LJ, Schreiber RD. Cancer immunoediting: from immunosurveillance to tumor escape. Nat Immunol. 2002 Nov;3(11):991-8.
3. Teng MW, Swann JB, Koebel CM, Schreiber RD, Smyth MJ. Immune-mediated dormancy: an equilibrium with cancer. J Leukoc Biol. 2008 Oct;84(4):988-93.
4. Ruffell B, DeNardo DG, Affara NI, Coussens LM. Lymphocytes in cancer development: polarization towards pro-tumor immunity. Cytokine Growth Factor Rev. 2010 Feb;21(1):3-10.
5. Kim R, Emi M, Tanabe K. Cancer immunoediting from immune surveillance to immune escape. Immunology. 2007 May;121(1):1-14.
6. Dunn GP, Old LJ, Schreiber RD. The three Es of cancer immunoediting. Annu Rev Immunol. 2004;22:329-60.
7. Schreiber RD, Old LJ, Smyth MJ. Cancer immunoediting: integrating immunity's roles in cancer suppression and promotion. Science. 2011 Mar 25;331(6024):1565-70.
8. Shankaran V, Ikeda H, Bruce AT, White JM, Swanson PE, Old LJ, Schreiber RD. IFNgamma and lymphocytes prevent primary tumour development and shape tumour immunogenicity. Nature. 2001 Apr 26;410(6832):1107-11.
9. Ganapathy V, Thangaraju M, Prasad PD. Nutrient transporters in cancer: relevance to Warburg hypothesis and beyond. Pharmacol Ther. 2009 Jan;121(1):29-40.
10. Morrison SJ, Kimble J. Asymmetric and symmetric stem-cell divisions in development and cancer. Nature. 2006 Jun 29;441(7097):1068-74.
11. O'Hare T, Corbin AS, Druker BJ. Targeted CML therapy: controlling drug resistance, seeking cure. Curr Opin Genet Dev. 2006 Feb;16(1):92-9.
12. Oravecz-Wilson KI, Philips ST, Yilmaz OH, Ames HM, Li L, Crawford BD, Gauvin AM, Lucas PC, Sitwala K, Downing JR, Morrison SJ, Ross TS. Persistence of leukemia-initiating cells in a conditional knockin model of an imatinib-responsive myeloproliferative disorder. Cancer Cell. 2009 Aug 4;16(2):137-48.
13. Li X, Lewis MT, Huang J, Gutierrez C, Osborne CK, Wu MF, Hilsenbeck SG, Pavlick A, Zhang X, Chamness GC, Wong H, Rosen J, Chang JC. Intrinsic resistance of tumorigenic breast cancer cells to chemotherapy. J Natl Cancer Inst. 2008 May 7;100(9):672-9.
14. Diehn M, Cho RW, Lobo NA, Kalisky T, Dorie MJ, Kulp AN, Qian D, Lam JS, Ailles LE, Wong M, Joshua B, Kaplan MJ, Wapnir I, Dirbas FM, Somlo G, Garberoglio C, Paz B, Shen J, Lau SK, Quake SR, Brown JM, Weissman IL, Clarke MF. Association of reactive oxygen species levels and radioresistance in cancer stem cells. Nature. 2009 Apr 9;458(7239):780-3.

15. Aktipis CA, Maley CC, Pepper JW. Dispersal evolution in neoplasms: the role of disregulated metabolism in the evolution of cell motility. Cancer Prev Res (Phila). 2012 Feb;5(2):266-75.
16. Wilenski U, Rand W. An introduction to agent-based modeling: Modeling natural, social and engineered complex systems with NetLogo. Cambridge, MA: MIT Press; 2009.
17. Fox CJ, Hammerman PS, Thompson CB. Fuel feeds function: energy metabolism and the T-cell response. Nat Rev Immunol. 2005 Nov;5(11):844-52.
18. Buttgereit F, Brand MD. A hierarchy of ATP-consuming processes in mammalian cells. Biochem J. 1995 Nov 15;312 ( Pt 1):163-7.
19. Buttgereit F, Burmester GR, Brand MD. Bioenergetics of immune functions: fundamental and therapeutic aspects. Immunol Today. 2000 Apr;21(4):192-9.
20. Maciver NJ, Jacobs SR, Wieman HL, Wofford JA, Coloff JL, Rathmell JC. Glucose metabolism in lymphocytes is a regulated process with significant effects on immune cell function and survival. J Leukoc Biol. 2008 Oct;84(4):949-57.
21. Di Tomaso T, Mazzoleni S, Wang E, Sovena G, Clavenna D, Franzin A, Mortini P, Ferrone S, Doglioni C, Marincola FM, Galli R, Parmiani G, Maccalli C. Immunobiological characterization of cancer stem cells isolated from glioblastoma patients. Clin Cancer Res. 2010 Feb 1;16(3):800-13.
22. Chikamatsu K, Takahashi G, Sakakura K, Ferrone S, Masuyama K. Immunoregulatory properties of CD44+ cancer stem-like cells in squamous cell carcinoma of the head and neck. Head Neck. 2011 Feb;33(2):208-15.
23. Inoda S, Hirohashi Y, Torigoe T, Morita R, Takahashi A, Asanuma H, Nakatsugawa M, Nishizawa S, Tamura Y, Tsuruma T, Terui T, Kondo T, Ishitani K, Hasegawa T, Hirata K, Sato N. Cytotoxic T lymphocytes efficiently recognize human colon cancer stem-like cells. Am J Pathol. 2011 Apr;178(4):1805-13.
24. Pienta KJ, McGregor N, Axelrod R, Axelrod DE. Ecological therapy for cancer: defining tumors using an ecosystem paradigm suggests new opportunities for novel cancer treatments. Transl Oncol. 2008 Dec;1(4):158-64.
25. Breivik J. Don't stop for repairs in a war zone: Darwinian evolution unites genes and environment in cancer development. Proc Natl Acad Sci U S A. 2001 May 8;98(10):5379-81.
26. Greaves M, Maley CC. Clonal evolution in cancer. Nature. 2012 Jan 18;481(7381):306-13.
27. Nowell PC. The clonal evolution of tumor cell populations. Science. 1976 Oct 1;194(4260):23-8.
28. Calabrese P, Tavaré S, Shibata D. Pretumor progression: clonal evolution of human stem cell populations. Am J Pathol. 2004 Apr;164(4):1337-46.
29. Greaves M. Cancer stem cells: back to Darwin?. Semin Cancer Biol. 2010 Apr;20(2):65-70.
30. Kareva I. What can ecology teach us about cancer?. Transl Oncol. 2011 Oct;4(5):266-70.

31. Merlo LM, Pepper JW, Reid BJ, Maley CC. Cancer as an evolutionary and ecological process. Nat Rev Cancer. 2006 Dec;6(12):924-35.
32. Tomlinson IP. Game-theory models of interactions between tumour cells. Eur J Cancer. 1997 Aug;33(9):1495-500.
33. Basanta D, Deutsch A. A game theoretical perspective on the somatic evolution of cancer. In: Bellomo N, editor. Selected topics on cancer modeling: genesis, evolution, immune competition, therapy. Boston: Birkhauser; 2008.
34. Basanta D, Simon M, Hatzikirou H, Deutsch A. Evolutionary game theory elucidates the role of glycolysis in glioma progression and invasion. Cell Prolif. 2008 Dec;41(6):980-7.
35. Mansury Y, Diggory M, Deisboeck TS. Evolutionary game theory in an agent-based brain tumor model: exploring the 'Genotype-Phenotype' link. J Theor Biol. 2006 Jan 7;238(1):146-56.
36. Thomas VL, Joel BS. Evolutionary game theory, natural selection, and Darwinian dynamics. Cambridge: Cambridge University Press; 2005.
37. Gatenby RA, Vincent TL. An evolutionary model of carcinogenesis. Cancer Res. 2003 Oct 1;63(19):6212-20.
38. Gatenby RA, Maini PK. Mathematical oncology: cancer summed up. Nature. 2003 Jan 23;421(6921):321.
39. Kareva I. Prisoner's dilemma in cancer metabolism. PLoS One. 2011;6(12):e28576.
40. Enderling H, Hlatky L, Hahnfeldt P. Immunoediting: evidence of the multifaceted role of the immune system in self-metastatic tumor growth. Theor Biol Med Model. 2012 Jul 28;9:31.
41. Hillen T, Enderling H, Hahnfeldt P. The tumor growth paradox and immune system-mediated selection for cancer stem cells. Bull Math Biol. 2013 Jan;75(1):161-84.
42. Moncharmont C, Levy A, Gilormini M, Bertrand G, Chargari C, Alphonse G, Ardail D, Rodriguez-Lafrasse C, Magné N. Targeting a cornerstone of radiation resistance: cancer stem cell. Cancer Lett. 2012 Sep 28;322(2):139-47.
43. McCord AM, Jamal M, Williams ES, Camphausen K, Tofilon PJ. CD133+ glioblastoma stem-like cells are radiosensitive with a defective DNA damage response compared with established cell lines. Clin Cancer Res. 2009 Aug 15;15(16):5145-53.
44. Marcato P, Dean CA, Giacomantonio CA, Lee PW. If cancer stem cells are resistant to current therapies, what's next?. Future Oncol. 2009 Aug;5(6):747-50.
45. Rich JN. Cancer stem cells in radiation resistance. Cancer Res. 2007 Oct 1;67(19):8980-4.
46. Schatton T, Frank MH. Antitumor immunity and cancer stem cells. Ann N Y Acad Sci. 2009 Sep;1176:154-69.
47. Hirohashi Y, Torigoe T, Inoda S, Takahashi A, Morita R, Nishizawa S, Tamura Y, Suzuki H, Toyota M, Sato N. Immune response against tumor antigens expressed on human cancer stem-like cells/tumor-initiating cells. Immunotherapy. 2010 Mar;2(2):201-11.

48. Todaro M, D'Asaro M, Caccamo N, Iovino F, Francipane MG, Meraviglia S, Orlando V, La Mendola C, Gulotta G, Salerno A, Dieli F, Stassi G. Efficient killing of human colon cancer stem cells by gammadelta T lymphocytes. J Immunol. 2009 Jun 1;182(11):7287-96.
49. Heddleston JM, Li Z, McLendon RE, Hjelmeland AB, Rich JN. The hypoxic microenvironment maintains glioblastoma stem cells and promotes reprogramming towards a cancer stem cell phenotype. Cell Cycle. 2009 Oct 15;8(20):3274-84.
50. Lin Q, Yun Z. Impact of the hypoxic tumor microenvironment on the regulation of cancer stem cell characteristics. Cancer Biol Ther. 2010 Jun 15;9(12):949-56.
51. Doloff JC, Waxman DJ. VEGF receptor inhibitors block the ability of metronomically dosed cyclophosphamide to activate innate immunity-induced tumor regression. Cancer Res. 2012 Mar 1;72(5):1103-15.